\documentclass[12pt]{article}

\usepackage[a4paper,margin=1in]{geometry}
\usepackage{amsmath,amssymb,amsthm}
\usepackage{graphicx}
\usepackage[hidelinks]{hyperref}

\title{Measurement as Sheafification: Context, Logic, and Truth after Quantum Mechanics}

\author{%
Partha Ghose\\[0.25em]
\small Tagore Centre for Natural Sciences and Philosophy, Rabindra Tirtha, New Town, Kolkata 700156, India\\
\small \texttt{partha.ghose@gmail.com}\\
}

\date{}

\begin{document}
\maketitle

\begin{abstract}
Quantum measurement is commonly posed as a dynamical tension between linear Schr\"odinger evolution and an ad hoc collapse rule. I argue that the deeper conflict is logical: quantum theory is inherently contextual, whereas the classical tradition presupposes a single global, Boolean valuation. Building on Bohr's complementarity, the Einstein--Podolsky--Rosen argument and Bell's theorem, I recast locality and completeness as the existence of a global section of a presheaf of value assignments over the category of measurement contexts. The absence of global sections expresses the impossibility of context-independent description, and \v{C}ech cohomology measures the resulting obstruction. The internal logic of the presheaf topos is intuitionistic, and the Ghose--Patra seven-valued contextual logic is exhibited as a finite Heyting algebra capturing patterns of truth, falsity and indeterminacy across incompatible contexts. Classical physics corresponds to the sheaf case, where compatible local data glue and Boolean logic is effectively restored. Measurement is therefore reinterpreted as sheafification of presheaf-valued truth rather than as a physical breakdown of unitarity. Finally, a $\sigma$--$\lambda$ dynamics motivated by stochastic mechanics provides a continuous interpolation between strongly contextual and approximately classical regimes, dissolving the usual measurement paradoxes and apparent nonlocality as artefacts of an illegitimate demand for global truth.
\end{abstract}


\section{The Classical Pact Between Logic and Physics}

Classical mechanics admits a description in which physical quantities possess definite values simultaneously (like position and momentum) and independently of how they are measured. This feature aligns seamlessly with Boolean logic, where propositions are either true or false absolutely. Over time, this harmony hardened into a tacit pact: logic came to be regarded as universally valid and ontologically neutral, while physics merely instantiated it. The success of classical physics concealed the contingency of this arrangement. Logic appeared to precede and constrain physics, rather than to arise from the structure of physical description.

This harmony between classical physics and classical logic did not arise by
historical accident alone. It received its most influential philosophical
formulation in Kant's response to Newtonian science \cite{Rohlf2024}. Kant's central insight was
that the universal validity and necessity exhibited by Newtonian mechanics could
not be derived from experience alone. Instead, such necessity must arise from
the structures through which experience itself becomes possible.

According to Kant, space and time are not properties of things in themselves but
forms of sensible intuition, while fundamental concepts such as object,
causality, and substance are conditions under which phenomena can be thought at
all. The laws of Newtonian physics thus appear universal and exceptionless
because they describe a world already organised by these a priori forms.
Crucially for our purposes, this organisation presupposes that physical systems
possess determinate properties at each moment in time and that propositions
about them admit unambiguous truth values.

In this way, Newtonian physics and classical logic were brought together within a
single framework. Boolean logic, with its sharp distinction between true and
false, was not merely compatible with classical mechanics; it appeared to be
forced upon us by the very conditions of possible experience. The logical form of
physical description was thus elevated from a contingent feature of successful
theories to a seemingly universal necessity.

The enduring influence of this Kantian synthesis lies in its stability: once the
logical preconditions of Newtonian science were identified with the
preconditions of experience as such, the idea that alternative physical theories
might call for alternative logical frameworks became effectively unthinkable. It
is this deeply ingrained assumption---rather than any specific physical postulate---
that quantum mechanics ultimately calls into question.

It is worth noting that even within classical physics this Kantian synthesis
was not left entirely unchallenged. Einstein's theory of relativity already
undermined one of Kant's central assumptions: the a priori status of Euclidean
geometry. As Einstein himself emphasized in his \emph{Autobiographical Notes} \cite{Einstein1949},
the geometric structure of space is not fixed by the conditions of thought
alone, but is subject to empirical determination by physical theory.

Yet despite this profound revision, relativity theory remained classical in a
crucial sense. Physical states were still described globally by fields defined
on spacetime, their evolution governed by deterministic equations, and physical
propositions retained unambiguous truth values. What changed was not the logical
form of physical description, but its geometric backdrop. Boolean logic survived
relativity intact.

In hindsight, this illustrates an important distinction. Relativity revealed
that geometry need not be a priori, but it left untouched the deeper assumption
that physical reality admits a single, context-independent description. It is
this remaining assumption---preserved by Einstein---that quantum mechanics
would ultimately force us to abandon, much to Einstein's discomfort.

\section{Quantum Mechanics as a Crisis of Global Truth}

Quantum mechanics marks a decisive break with the classical assumption that
physical properties admit a single, context-independent description. This
departure is already implicit in Bohr's formulation of the Complementarity
Principle \cite{BohrRosenfeld1996}, long before later no-go theorems made the point mathematically
explicit. Bohr emphasized that the description of quantum phenomena depends
ineliminably on the experimental conditions under which they are observed. The
formalism itself assigns no meaning to properties in abstraction from such
conditions.

Bohr's use of the double-slit experiment to illustrate complementarity is
especially telling. Under one experimental arrangement, designed to register
interference, the quantum object must be described as a wave propagating through
both slits. Under a mutually exclusive arrangement, designed to determine which
slit the object passes through, the description is necessarily particle-like.
These two descriptions are not merely practically incompatible; the conditions
that make one meaningful physically preclude the other. There is no single
experimental context in which both descriptions can be jointly applied. What is
crucial is that each description is perfectly well-defined and objective within
its own context, yet no global synthesis of the two is available.

Although Bohr did not formulate his insights in the language of formal logic, the
implication is clear: the truth of propositions about quantum systems is
conditioned by the experimental context. The attempt to ascribe simultaneous
truth to all such propositions reproduces a classical demand that the theory
itself does not support. Complementarity thus introduces contextuality into
quantum mechanics at a foundational level, even without appealing to hidden
variables or metaphysical assumptions.

This contextuality was later sharpened and formalized through results such as
the Kochen--Specker theorem \cite{KochenSpecker1967}. Unlike Bell-type arguments \cite{Bell1964}, which rule out local
hidden-variable theories under assumptions of separability or locality, the
Kochen--Specker theorem demonstrates the impossibility of assigning definite
values to quantum observables in a way that is both non-contextual and consistent
with the functional relations between observables. No assignment of pre-existing
values can reproduce the predictions of quantum mechanics if those values are
required to be independent of measurement context.

The notion of contextuality that emerges here differs in form from Bohr's
original discussion, but not in substance. In both cases, the obstruction lies
in the attempt to maintain a globally valid assignment of truth values across
mutually incompatible experimental arrangements. What Kochen--Specker makes
explicit is that \emph{this obstruction is not a consequence of practical limitations
or incomplete knowledge, but a structural feature of the theory itself}.

Quantum mechanics thus confronts us with a new situation: propositions that are
perfectly meaningful and decidable within a given context cannot, in general, be
assembled into a single context-independent account. This failure of global
truth is not pathological; it is intrinsic to the quantum description of nature.
The persistence of the measurement problem and related paradoxes is a reflection
of the continued use of classical, Boolean logic in a domain where its basic
presuppositions no longer apply.

It is worth emphasising that several of the founders and early critics of
quantum mechanics came close to recognising this impasse, yet stopped just
short of its logical resolution. Bohr's doctrine of complementarity
correctly identified the indispensability of mutually incompatible
experimental contexts, but retained a classical conception of logical
description within each context without formalising the logical
relations between them. The Kochen--Specker theorem went further in showing
that no global assignment of definite values \emph{is} possible, but it did so as a
negative result: it demonstrated the impossibility of a classical valuation
without replacing it by a new, explicitly contextual logic.

Von Neumann and Birkhoff made the boldest attempt to reform logic itself by
introducing quantum logic, replacing Boolean lattices by the lattice of
closed subspaces of Hilbert space \cite{BirkhoffvonNeumann1936}. However, 
this approach preserved a global,
context-independent logical structure at a deeper level, and thereby shifted
rather than resolved the underlying tension. Logical operations were altered,
but the assumption of a single overarching logical space remained intact.

Reichenbach's proposal of a three-valued probability logic likewise
acknowledged the inadequacy of classical truth values, yet interpreted
indeterminacy probabilistically rather than contextually \cite{Reichenbach1944}. As a result, it
treated indefiniteness as a matter of partial truth or ignorance, rather than
as a structural feature arising from incompatible descriptions.

What these influential approaches share is a reluctance to let logic itself
be conditioned by physical context. Quantum mechanics, however, persistently
forces such a conditioning. The ambiguity that remains in the measurement
problem is not accidental; it is the residue of an unmet demand to rethink
the very notion of truth in a context-dependent way. This essay proceeds from
the claim that only by doing so can the conceptual tension at the foundation
of quantum mechanics finally be dissolved.

\subsection*{The EPR Argument and the Demand for Global Truth}

The tension between quantum mechanics and classical logic was brought
into sharp focus by the celebrated Einstein--Podolsky--Rosen (EPR)
argument \cite{EinsteinPodolskyRosen1935}.  EPR considered two
systems that have interacted and then separated, so that their joint
quantum state displays perfect correlations between suitably chosen
observables.  If one measures an observable $A$ on the first system, one
can predict with certainty the outcome of a corresponding observable $B$
on the distant system, without in any way disturbing it.  EPR then
introduced a criterion of reality: if, without disturbing a system, one
can predict with certainty the value of a physical quantity, then there
exists an element of physical reality corresponding to that quantity.

Assuming locality in the strict sense that operations on the first
system cannot instantaneously affect the second, EPR concluded that the
second system must possess definite values for both of two
incompatible observables, depending on which measurement is performed
on the first.  Since the quantum formalism assigns no such joint values
to incompatible observables, EPR inferred that the quantum-mechanical
description of physical reality is incomplete.  In effect, their
argument demands that, under the joint assumptions of locality and
perfect correlation, there must exist a single, global assignment of
values to all relevant observables: a global truth function extending
across all measurement contexts.

Bohr's immediate reply to EPR in 1935 made this contextual dependence
completely explicit \cite{Bohr1935}. He grants that the two
particles form a single entangled whole and that a measurement on one
side allows one to predict with certainty the outcome of a suitably
chosen measurement on the distant partner.  But he denies that this
justifies ascribing \emph{simultaneous} reality to noncommuting
quantities of the distant particle.  The crucial point, for Bohr, is
that the very meaning of ``position of the second particle'' or
``momentum of the second particle'' is fixed only within a
well-defined experimental arrangement.  The EPR argument illicitly
combines facts obtained under mutually exclusive arrangements into a
single global description.  In Bohr's language, there is indeed ``no
mechanical disturbance'' of the distant system; rather, the choice of
measurement here changes the \emph{conditions which define the
possible predictions} there.  In modern terms, Bohr is already
arguing that truth for quantum propositions is irreducibly
\emph{contextual}: one cannot form a single Boolean algebra of
properties spanning all complementary setups at once, and the EPR
demand for such a global truth assignment is therefore conceptually
misplaced rather than empirically refuted.
In contemporary language, this is precisely the demand that there should
exist a global section of the presheaf of value assignments.  EPR took
the existence of such a global section to be a requirement of locality
and completeness; the subsequent development of Bell's theorem showed
that no local hidden-variable theory can reproduce the full range of
quantum correlations \cite{Bell1964}.  The usual conclusion is that one must
either give up locality or accept some form of nonlocal ``spooky
action-at-a-distance.''

The perspective adopted in this paper is different.  The EPR argument
reveals not a failure of quantum mechanics to supply missing variables,
but a clash between the presheaf-like, contextual structure enforced by
the theory and the assumption that there must nonetheless exist a single
global valuation compatible with locality.  Once one acknowledges that
quantum truth is intrinsically context-dependent and that the relevant
semantic object is a presheaf without global section, the EPR demand for
a local global truth assignment is seen to be ill-posed.  The
``nonlocality'' that appears in the standard reading of Bell's theorem
is thus a symptom of insisting on a global logical structure that the
physical theory does not support, rather than direct evidence for
superluminal influences.

This way of phrasing the EPR and Bell arguments is closely related to
the sheaf-theoretic analysis of non-locality and contextuality developed
by Abramsky and Brandenburger \cite{AbramskyBrandenburger2011}, in which contextuality is likewise
characterised as the obstruction to the existence of global sections of
a presheaf of outcome assignments.

\section{Why the Measurement Problem Is Conceptually Misplaced}

Once the impossibility of global truth assignments has been recognised, the
traditional framing of the measurement problem comes into focus as a
misdiagnosis. The problem is conventionally posed as a demand for a physical
mechanism that transforms quantum superpositions into definite classical
outcomes. This demand, however, already presupposes that physical reality must
at all times admit a single, global, context-independent description. It is
this presupposition---rather than any deficiency in the quantum formalism---that
gives rise to the appearance of paradox.

Classical logic embodies the assumption that propositions possess definite
truth values independently of the conditions under which they are evaluated.
In classical physics, this assumption is unproblematic: the theory admits a
global phase space in which all observables have simultaneous, well-defined
values. Logical operations such as conjunction and disjunction merely mirror
this underlying structure. When classical logic is carried over uncritically
into quantum theory, however, it enforces a requirement that the theory itself
systematically refuses to satisfy. The insistence on global truth becomes an
external constraint imposed on a fundamentally contextual framework.

The resulting conflict inevitably manifests as a physical catastrophe. If the
quantum state is taken to evolve unitarily at all times, classical logic
demands that measurement outcomes nevertheless be globally definite. This
forces the introduction of an additional dynamical process---collapse---whose sole
function is to restore a logical condition that the unitary dynamics violates.
Collapse thus appears not as a physical necessity arising from the theory, but
as a logical repair mechanism introduced to save a classical notion of truth.

 This tension was already sharply perceived by Schr\"{o}dinger in his 1935 analysis of the 
 `present situation in quantum mechanics' \cite{Schrodinger1935}. 
Starting from the linear, deterministic Schr\"{o}dinger evolution, he
stresses that, if the formalism is taken at face value, microscopic
superpositions are unavoidably amplified into grotesque macroscopic
superpositions---live cat plus dead cat.  At the same time, he notes
that the standard account prescribes a \emph{suspension} of this
unitary evolution during measurement, replacing it by an abrupt,
probabilistic ``jump'' to a definite outcome, without offering any
precise dynamical law for when and how this happens.  Schr{\"o}dinger
thus treats the collapse postulate less as a physical mechanism than
as a sign of conceptual deficiency: a rule introduced purely to
reconcile the formalism with the definite character of experience.
In the sheaf-theoretic reading proposed here, this tension is not
resolved by modifying the Schr{\"o}dinger dynamics, but by recognising
that the passage from entangled, context-dependent descriptions to
classically communicable outcomes is a change in the \emph{logical}
regime---a sheafification of presheaf-valued truth---rather than a
mysterious violation of the unitary law.

Seen in this light, the familiar paradoxes of quantum mechanics acquire a new
interpretation. Schr\"odinger's cat, Wigner's friend, and related scenarios do
not signal a breakdown of physical law; they reveal the incompatibility between
a contextual theory and a non-contextual logic. Each paradox arises when one
attempts to combine descriptions that are valid within different contexts into
a single global account. The contradiction is logical before it is physical.

Interpretational strategies respond to this tension in characteristic ways.
Some, such as objective collapse models, modify the dynamics to enforce global
definiteness. Others, such as many-worlds approaches, preserve unitary evolution
at the cost of multiplying classical realities. Still others introduce
privileged observers or appeal to consciousness. What these diverse responses
have in common is their attempt to secure classical logical absoluteness rather
than to question it. They all treat context-dependence as an anomaly to be
eliminated, rather than as a structural feature to be accommodated.

The central claim of this essay is that this strategy is misguided. Quantum
mechanics does not demand the explanation of a mysterious physical transition,
but the abandonment of an inappropriate logical expectation. Once truth itself
is recognised as context-dependent, the demand for a universal, collapse-inducing
mechanism loses its force. The measurement problem, properly understood, is not
a problem that calls for a new physical process, but a symptom of insisting on a
form of logic that quantum theory no longer supports.

It is worth noting in this connection that Einstein complained in a letter written on 19 June, 1935 to 
Schr\"{o}dinger that the EPR paper, drafted by Podolsky ``for reasons of language'' had allowed the main point to be ``smothered by formalism (Gelehrsamkeit)'' \cite{Fine1996}. In his later reflections, he distanced
his concern from specific hidden-variable programmes and suggested instead that
the difficulty lay deeper than the addition of further physical parameters. It
is therefore clear that Einstein felt profoundly that something essential
was missing in quantum mechanics, even though he did not---and perhaps could not have---
anticipated that the missing element would lie in the logical structure
through which the theory is interpreted.

\section{Context as the Primitive: From Measurement Setups to Categories}
If the measurement problem is really a problem of logic rather than of
dynamics, we must ask what the basic logical units are.  In the present
proposal the primitive notion is that of a \emph{measurement context}.
Roughly speaking, a context is a physically real experimental
arrangement: a choice of observables, an arrangement of apparatus, a
temporal ordering of interventions, and an environment in which the
standard quantum predictions can be applied without further
qualification.

Two features of such contexts are crucial.  First, within any given
context, the usual quantum formalism delivers definite probabilistic
predictions; once an outcome is registered, it can be reported as an
unambiguous fact.  Second, different contexts may be \emph{mutually
incompatible}: the conditions that make one arrangement meaningful may
exclude another.  Bohr already insisted on this point in his discussions
of complementary experiments; the double-slit arrangement designed to
reveal interference excludes, as a matter of physical principle, the
arrangement that would reveal which-path information.  There is no
meta-context in which both descriptions apply simultaneously.

Mathematically, it is natural to organise such contexts into a category
$C$.  The \emph{objects} of $C$ are measurement contexts.  A morphism
$f : C \to D$ is interpreted as a physically meaningful refinement or
coarse-graining: passing from $C$ to $D$ by adding compatible
observables, increasing resolution, or otherwise sharpening the
description.  Composition corresponds to performing such refinements in
succession, and identity arrows represent leaving a context unchanged.
Importantly, \emph{not} every pair of contexts need be related by a
morphism: the absence of an arrow $C_1 \to C_2$ records physical
incompatibility rather than logical contradiction.

Within each context $C$ one may consider the propositions about the
system that are testable in $C$ and the corresponding truth values
assigned by the theory.  Classically this data would be packaged into a
single global phase space; in the present viewpoint it is deliberately
kept \emph{local} to each context.  The question of how such local
assignments behave when one passes from one context to another, and
whether they can be glued into a single global description, is then
expressed in the language of functors on $C$.

This is precisely where presheaves enter.  A presheaf on $C$ assigns to
each context $C$ the appropriate set of truth assignments, expectation
values, or outcome structures, together with restriction maps along
morphisms in $C$.  The next section will show how presheaves provide the
natural semantic framework for context-dependent quantum truth, and how
the failure of a global section expresses, in a mathematically precise
way, the breakdown of classical, context-independent description.

\section{Presheaves as the Natural Semantics of Quantum Truth}

The conceptual shift required by quantum mechanics is most clearly expressed
once attention is moved away from absolute states and toward relations between
contexts. Category theory \cite{MacLane1998} (see also Appendix A for a concise mathematical introduction)
 provides a language precisely suited for this purpose,
as it allows one to describe not only objects, but also the web of relations
between them. Related categorical reconstructions of quantum mechanics have been
developed by Abramsky and Coecke \cite{AbramskyCoecke2004,AbramskyCoecke2008}, who axiomatise the
theory at the level of strongly compact closed (or dagger compact)
categories in order to capture the compositional structure of quantum
processes and information flow.  Their ``categorical quantum mechanics''
programme shows that a large part of the Hilbert-space formalism can be
recovered from purely structural assumptions on such process categories.
The use of category theory in the present work is complementary to this:
rather than reaxiomatising the state--process calculus itself, it
organises measurement contexts into a category and uses presheaves and
cohomological tools to analyse the logical and contextual structure of
truth and measurement. In the present setting, the objects of interest are measurement
contexts, understood as physically real experimental arrangements, while the
relations describe how one context may refine or extend another. Importantly,
not all contexts need be mutually related: incompatibility is expressed not by
contradiction, but by the absence of a relation.

A category, in this intuitive sense, is nothing more than a structured
collection of contexts together with the physically meaningful ways of passing
from one to another. What matters is not the internal composition of each
context in isolation, but how descriptions change when the context is varied.
This relational viewpoint already departs from classical thinking, where a
single global domain is presupposed from the outset.

Presheaves arise naturally once one asks how truth or physical description is
assigned relative to contexts. A presheaf associates to each context the set of
statements, values, or assignments that are meaningful within that context, and
specifies how these assignments are restricted when passing to a more limited
or coarser context. Crucially, presheaves make no demand that locally valid
descriptions fit together into a single global picture. They formalise the
possibility that truth may be well-defined within each context, yet resist
unification across incompatible ones.

This feature makes presheaves particularly well suited to quantum mechanics.
The theory consistently provides context-dependent descriptions---expectation
values, probabilities, or definite outcomes---while denying the existence of a
single context in which all such descriptions can be simultaneously realised.
The failure of global truth in quantum mechanics is therefore not an anomaly,
but exactly the behaviour presheaves are designed to accommodate.

Sheaves represent a special case within this framework. A sheaf imposes a
stronger consistency requirement: whenever local descriptions agree on all
overlaps between contexts, they must arise from a unique global description.
This is precisely the logical structure tacitly assumed in classical physics.
From this perspective, classical logic emerges not as a fundamental principle,
but as a consequence of working in regimes where the sheaf condition happens to
be satisfied.

The distinction between presheaves and sheaves thus mirrors the distinction
between quantum and classical truth. Quantum theory naturally gives rise to
presheaf-like semantics, while classical physics corresponds to the special case
in which presheaf data can be glued into global sections.

Cohomology provides a further conceptual refinement. Rather than merely stating
that global truth may fail, cohomological tools offer a way to characterise and
measure this failure. When locally valid assignments cannot be consistently
glued together, the obstruction can be represented by cohomology classes. In
physical terms, nontrivial cohomology signals the presence of irreducible
contextuality, while the vanishing of such obstructions corresponds to the
emergence of classical, globally consistent descriptions.

In this way, categories encode contexts, presheaves encode context-dependent
truth, sheaves encode classical consistency, and cohomology records the precise
sense in which quantum descriptions resist global unification. These ideas will
provide the conceptual foundation for the more detailed analysis that follows.

\subsection*{Intuitionistic Logic Inside Presheaves}

The logical behaviour of presheaves already departs from the classical,
two-valued picture in a fundamental way. The internal logic of any category
of presheaves is not Boolean but \emph{intuitionistic}. In classical logic,
propositions are assumed to be either true or false absolutely, and the
principles of excluded middle ($P \vee \neg P$) and double negation
($\neg \neg P \Rightarrow P$) hold universally. Intuitionistic logic
abandons these principles as general laws. A statement is taken to be
true only when a suitable construction or justification is available, and
it need not be the case that either $P$ or $\neg P$ holds in the absence
of such a construction.

Presheaves give a natural semantic home for intuitionistic logic because
truth is evaluated \emph{relative to context}. A proposition about a
quantum system may have a well-defined truth value in one measurement
context, a different value in another, and no determinate value at all
in incompatible contexts. The collection of all such truth values in a
presheaf topos forms a Heyting algebra rather than a Boolean algebra: it
supports conjunction, disjunction and implication, but not in a way that
forces every proposition to be either globally true or globally false.
Intuitively, truth can grow monotonically as contexts are refined, and the
logical structure records this possibility.

Seen in this light, classical Boolean logic appears as a special,
degenerate case of intuitionistic logic, recovered precisely when the
presheaf data collapse to a single global section. In that regime the
Heyting algebra of truth values becomes Boolean: excluded middle and
double negation are restored, and propositions behave as if they had
context-independent truth values. Outside this regime, however, the more
flexible, context-sensitive structure of intuitionistic logic is
indispensable.

\subsection*{Contextual Multi-Valued Logic}

This intuitionistic background naturally paves the way for a
multi-valued treatment of quantum propositions. Once truth is recognised
as context-dependent, it is no longer adequate to speak only of ``true''
and ``false'' simpliciter. One must also distinguish cases in which a
proposition is true in one context and false in another, or true in one
context and indeterminate in another, and so on. These patterns of
contextual variation themselves become the relevant ``truth values.''

The seven-valued scheme proposed by Ghose and Patra \cite{GhosePatra2024} can be viewed in
precisely this way (see Appendix B for some technical details). Each of the seven values---``true,'' ``false,''
``indeterminate,'' and the four mixed cases---represents a distinct mode
in which truth may vary across incompatible contexts. Formally, these
seven values can be organised into a finite Heyting algebra: they admit
logical operations of conjunction, disjunction and implication, but do
not collapse to a simple two-valued Boolean structure. In this sense, the
GP logic may be regarded as a concrete, context-sensitive instance of
intuitionistic semantics tailored to quantum phenomena.

What distinguishes the GP scheme from more abstract versions of
intuitionistic logic is its explicitly contextual and relational
character. The multi-valuedness does not arise from a vague notion of
partial truth or degree of belief, but from the way propositions about
quantum systems take on different truth values in physically incompatible
measurement arrangements. The resulting structure is therefore not a
mere generalisation of classical logic, but a logical reflection of the
contextual architecture that presheaves and their intuitionistic logic
make precise.

\section{Sheaves and the Emergence of Classical Logic}

A sheaf is a presheaf with the added property that compatible local data uniquely determine a global section. Sheaves thus encode the logical structure presupposed by classical physics: context-independent truth and Boolean reasoning. Classical logic is not fundamental in this view; it emerges when physical circumstances permit gluing. Classicality is therefore a special regime, characterised by the satisfaction of sheaf conditions.

In the next subsection, this emergent view will be applied directly to the measurement process.

\subsubsection*{Measurement Reinterpreted: Sheafification, Not Physical Collapse}

From this perspective, the traditional problem of measurement acquires a very
different character. In the interpretation associated with Bohr, there is no
measurement problem in the later, technical sense. Bohr consistently rejected
the demand for a physical account of wavefunction collapse and instead
emphasised the necessity of describing experimental outcomes in a classical
language refined through the development of Newtonian mechanics and Maxwellian
electrodynamics. What mattered for him was not the ontological status of the
quantum state, but the possibility of unambiguous communication of experimental
results.

This insistence on classical description may be understood, in modern terms, as
a demand for global consistency. Classical language presupposes that outcomes
can be reported as definite, shared facts, independent of the particular
experimental context in which they are obtained. In the present framework,
this corresponds precisely to the existence of global sections---descriptions
that are valid across all relevant contexts and can therefore be communicated
without ambiguity. Bohr's requirement thus amounts to privileging those
descriptions that satisfy a sheaf-like condition.

What Bohr carefully avoided was the attribution of such global descriptions to
quantum systems prior to measurement. Quantum theory, in his view, provides only
context-dependent accounts tied to specific experimental arrangements.
Measurement does not reveal a pre-existing global state; it marks the point at
which a description must be rendered in classical terms in order to be
communicable. In this sense, Bohr's ``cut'' separates presheaf-level quantum
descriptions from their sheaf-level classical articulation.

Reinterpreted in this way, measurement need not be understood as a physical
collapse interrupting unitary dynamics. Rather, it is a logical transition from
contextual, presheaf-based descriptions to globally consistent, sheaf-like ones.
The apparent discontinuity arises not in the underlying dynamics, but in the
imposition of a requirement of global truth. The collapse postulate enters only
when this logical transition is misread as a physical process.

Seen through the lens of sheaf theory, Bohr's insistence on classical language
appears not as an ad hoc philosophical restriction, but as an implicit
recognition that communication itself requires globalisation. The
measurement problem emerges only when this requirement is demanded of quantum
descriptions themselves, rather than of the language in which their outcomes
are reported.

\section{Dynamics as a Continuous Passage Between Logical Regimes}

Up to this point the discussion has been largely structural: quantum
theory was seen to enforce a presheaf-like, context-dependent notion of
truth, while classical physics corresponded to the special case in which
those presheaf data can be glued into a sheaf of globally valid
descriptions. To make this picture physically credible, one needs a
dynamical mechanism that allows for a \emph{continuous} passage between
these two regimes. Nelson's stochastic mechanics provides a natural
starting point \cite{Nelson1966}.

Nelson begins from the classical Newtonian picture, but assumes that
particles undergo a Brownian motion with a diffusion constant $\sigma$
superposed on their regular motion. Instead of a single velocity field,
one has forward and backward mean velocities, and the particle
trajectories are described by stochastic differential equations. The key
result is that, under suitable assumptions, the ensemble dynamics of such
stochastic trajectories can be recast into the familiar Schr\"odinger
equation. The Planck constant $\hbar$ then appears not as a primitive
quantum postulate, but as a parameter fixed by the relation
\[
\hbar = m \sigma,
\]
where $m$ is the mass and $\sigma$ is the diffusion parameter characterising the underlying Brownian motion.\footnote{More standardly one writes
$\nu = \hbar/(2m)$ for the diffusion constant; the present notation
absorbs numerical factors into $\sigma$ for simplicity.}

This identification has an important conceptual consequence. It shows
that what is usually regarded as a fundamental constant can be viewed, at
the level of the underlying stochastic processes, as a compound parameter linking mass
and diffusion. In particular, one may consider variations of $m$ and
$\sigma$ that leave $\hbar$ fixed, or, more generally, explore regimes in
which the effective strength of the underlying diffusion is altered. The
degree to which the resulting dynamics exhibits characteristic quantum
features can then be controlled continuously.

A complementary way to express this continuous control is to work with
the hydrodynamic form of the Schr\"odinger equation. Writing the wave
function in polar form,
\[
\psi = \sqrt{\rho}\, e^{iS/\hbar},
\]
and separating real and imaginary parts, one obtains a continuity equation
for the probability density $\rho$ and a modified Hamilton--Jacobi equation
for the phase $S$:
\[
\frac{\partial S}{\partial t}
+ \frac{(\nabla S)^2}{2m} + V + Q = 0,
\]
where $V$ is the classical potential and
\[
Q = -\,\frac{\hbar^2}{2m}\,\frac{\nabla^2 \sqrt{\rho}}{\sqrt{\rho}}
\]
is the so-called quantum potential. Setting $Q=0$ recovers the classical
Hamilton--Jacobi equation; retaining $Q$ yields the full quantum dynamics.

Rosen \cite{Rosen1964} observed that the classical Hamilton--Jacobi and continuity
equations can themselves be combined into a complex ``classical
Schr\"odinger equation'' in which the quantum potential term is absent.
From this viewpoint, the essential difference between classical and quantum
dynamics is precisely the presence or absence of $Q$. This suggests
introducing a one-parameter family of interpolating dynamics by writing
\[
\frac{\partial S}{\partial t}
+ \frac{(\nabla S)^2}{2m} + V + \lambda\, Q = 0,
\]
with $0 \leq \lambda \leq 1$ \cite{Ghose2002}. The case $\lambda = 1$ reproduces the
usual quantum dynamics, while $\lambda = 0$ yields the classical
Hamilton--Jacobi equation. Intermediate values of $\lambda$ describe
regimes in which the influence of the quantum potential is progressively
suppressed, and the dynamics moves continuously from quantum-like to
classical-like behaviour.

In the stochastic picture, the parameter $\lambda$ can be regarded as a
function of the underlying diffusion, $\lambda = \lambda(\sigma)$,
encoding how strongly the quantum potential survives coarse-graining and
environmental influence. Large diffusion (or suitable choices of $m$ and
$\sigma$ satisfying $\hbar = m\sigma$) support $\lambda \approx 1$ and
hence fully quantum behaviour; as the effective diffusion weakens, the
quantum potential term is diminished and $\lambda$ tends toward $0$. The
$\sigma$--$\lambda$ dynamics thus implements a \emph{continuous
deformation} of the underlying Hamilton--Jacobi structure between quantum
and classical limits.

Crucially, this continuous dynamical interpolation can be read as a
continuous passage between logical regimes. In the $\lambda \approx 1$
regime, where the quantum potential is fully active, trajectories are
highly sensitive to contextual information, and the space of measurement
contexts exhibits strong nontriviality. This is the domain in which
presheaf semantics and intuitionistic, multi-valued logic are indispensable,
and cohomological obstructions to global truth are generically present. As
$\lambda$ decreases and the quantum potential is progressively suppressed,
trajectories approach classical behaviour, incompatibilities between
contexts lose their operational significance, and the presheaf of truth
values becomes increasingly close to a sheaf. In the limit $\lambda \to 0$,
the cohomological obstructions vanish, a global section emerges, and
classical Boolean logic is effectively restored.

In this way, the $\sigma$--$\lambda$ dynamics does more than interpolate
between two mathematical equations, or even between two physical regimes.
It realises a continuous transition from a world in which truth is
irreducibly contextual and presheaf-based, to one in which truth can be
treated, to excellent approximation, as global and classical. Measurement,
understood as the selection of a sheaf-like description, is thereby
anchored in a dynamical process rather than in a postulated discontinuity.

\section{Logic as Emergent and the Dissolution of the Measurement Problem}

We are now in a position to look back and reconsider what has been at
stake in the measurement problem. The traditional formulation takes for
granted that there is a single, context-independent logical framework in
which all physical propositions must be evaluated. Within that framework,
the coexistence of unitary evolution and collapse appears paradoxical.
Unitary dynamics spreads possibilities into superpositions; collapse
restores definiteness in a way that seems to violate the very principles
governing the evolution between measurements. The problem, as usually
stated, is how to reconcile these two modes of description within one
physical theory.

The route taken in this essay is different from the standard treatments.
Rather than starting from the quantum formalism and asking how it might be
made to fit within a fixed classical logical framework, we have asked what
kind of logical framework the formalism itself suggests. In the perspective
developed here, quantum mechanics does not undermine the idea of a coherent
physical world, but it does challenge the assumption that this world must
always admit a single, context-independent global description. The category
of measurement contexts, and the presheaves defined over it, provide a
mathematically precise language for articulating this re-interpretation. Quantum
theory furnishes context-dependent descriptions that are internally
consistent within each context, yet resist amalgamation into a single
global picture. Classical physics, by contrast, corresponds to the
special case in which these presheaf data satisfy a sheaf condition,
admitting global sections and thereby supporting a Boolean notion of
truth.

From this perspective, \emph{logic is no longer a fixed background against
which physics unfolds, but an emergent structure reflecting the way
physical contexts are organised}. In strongly quantum regimes, where
contextuality is ineliminable and cohomological obstructions to global
truth are present, the appropriate internal logic is intuitionistic and
multi-valued, exemplified by context-sensitive schemes such as the
GP seven-valued logic. In classical regimes, where the
obstructions vanish and global sections exist, this richer logical
structure collapses to the familiar two-valued Boolean form. \emph{Logic, in
other words, is conditioned by the physical organisation of contexts}:
it is presheaf-like and contextual when the world forces contextuality
upon us, and effectively Boolean when the world allows globalisation.

The $\sigma$--$\lambda$ dynamics gives this picture a continuous
physical realisation. By interpreting the quantum potential as a
$\lambda$-scaled term interpolating between quantum and classical
Hamilton--Jacobi dynamics, and linking $\lambda$ to the underlying
stochastic parameter $\sigma$, one obtains a smooth passage from a
regime in which contextual effects are dominant to one in which they
are negligible. As $\lambda$ moves from $1$ toward $0$, the influence
of the quantum potential fades, the operational distinction between
incompatible contexts diminishes, and the presheaf of truth values
becomes increasingly close to a sheaf. In the limit, cohomological
obstructions disappear and a global, approximately classical description
emerges. The transition from quantum to classical is thus not a sudden
jump, but a deformation of both dynamics and logic.

In this light, the measurement problem loses its air of mystery. What
had appeared as a physical catastrophe---the abrupt collapse of the
wavefunction---can now be understood as a change in the admissible form
of description. \emph{Measurement corresponds to the selection, in practice,
of sheaf-like, globally communicable accounts from an underlying
presheaf of contextual quantum possibilities}. Collapse is not a
fundamental physical process layered on top of unitary evolution, but a
logical projection associated with the demand for classical,
context-independent reports of experimental outcomes. The paradox arises
only if one insists that the presheaf-level quantum description must
itself obey classical logical constraints.

Thus the measurement problem is not so much solved as dissolved. Once
logic is recognised as emergent from the physical organisation of
contexts, rather than imposed upon it, there is no longer a need to
postulate an ad hoc collapse mechanism or to invoke observers or
consciousness as external agents. The two ``processes'' of von Neumann
are reinterpreted as two regimes of description: a presheaf-like,
contextual regime appropriate to quantum systems, and a sheaf-like,
Boolean regime appropriate to classical communication and macroscopic
experience. The task is not to reconcile incompatible dynamics within a
fixed logical frame, but to understand how different logical structures
arise from the same underlying physical theory.

\section{A Historical Afterword: Contextual Logic and Early Anticipations}

The view developed in this essay has been motivated entirely by the
internal demands of quantum theory: the breakdown of global truth, the
centrality of measurement contexts, and the need for a logical framework
in which context-dependence is not an anomaly but a structural feature.
It is therefore striking, though historically accidental, that closely
related ideas appear in a very different setting within classical Indian
philosophy.

The Jaina doctrine of sevenfold predication, or \emph{Saptabha\.{n}g\={\i}}, is
often presented as a refinement of the broader principle of
\emph{sy\={a}dv\={a}da} (the ``doctrine of may-be'') \cite{Burch1964}. Instead of assigning a
single, context-independent truth value to a proposition, the Jaina
logicians distinguished seven possible predications: ``in some respect,
it is''; ``in some respect, it is not''; ``in some respect, it both is
and is not''; ``in some respect, it is indescribable''; and three
further mixed combinations involving indescribability. The crucial point
is that each predication is explicitly indexed by the qualifier
\emph{sy\={a}t} (``in some respect''): truth and falsity are not taken as
absolute, but as relative to a standpoint, viewpoint, or mode of
consideration.

In its original setting, this multi-valued, standpoint-dependent logic
served primarily metaphysical and epistemological aims. It was intended
to reconcile apparently incompatible perspectives on a complex reality
without collapsing them into a single, privileged view. There is no
suggestion of Hilbert spaces, measurement operators, or stochastic
dynamics. Nevertheless, the formal pattern is recognisably that of a
contextual logic: the same proposition may receive different evaluations
under different, mutually irreducible perspectives, and the task of
logic is to classify these evaluative patterns rather than to eliminate
them.

The contextual seven-valued scheme developed by Ghose and Patra
transposes this basic idea into a quantum setting. The seven values are
no longer vague modalities of assertion, but precise patterns of truth,
falsity, and indeterminacy across incompatible measurement contexts.
Formally, as argued earlier, they can be represented as subobjects in a
presheaf topos over the category of contexts, and their structure is
that of a finite Heyting algebra rather than a Boolean one. In this
sense, the GP logic may be regarded as a concrete, physically motivated
instance of the kind of intuitionistic, context-sensitive semantics that
presheaves naturally support.

It would be a mistake, however, to present the Jaina tradition as a
``source'' of quantum logic, or to suggest that quantum mechanics has
been anticipated in any straightforward way by Saptabha\.{n}g\={\i}. The
historical development of quantum theory is independent, and the
mathematical tools employed here are entirely modern. The point of
recalling the Jaina doctrine is more modest and, perhaps, more
interesting: it shows that the idea of a reality that admits only
context-dependent descriptions, and of a logic that classifies such
descriptions without forcing them into a single global frame, is not an
alien intrusion into human thought. Quantum theory compels us, for
strictly physical reasons, to rediscover a possibility that had already
been explored, in another guise, in a very different context and in a very different intellectual
tradition.

\section{Summary and Outlook}

The analysis developed in this paper has suggested that the
measurement problem in quantum mechanics is not primarily a dynamical
paradox, but a manifestation of a deeper logical tension.  The
standard formulation tacitly presupposes that physical reality must be
describable by a single, context-independent, globally valid Boolean
valuation.  Quantum theory, by contrast, forces upon us a world in
which experimental arrangements are mutually incompatible and in which
no global assignment of sharp values to all observables is possible.
The apparent conflict between unitary evolution and wavefunction
collapse, and the familiar worries about nonlocal ``spooky
action-at-a-distance'', are symptoms of this mismatch between a
classical logical ideal and the contextual structure imposed by the
quantum formalism itself.

By making measurement contexts explicit and organising them into a
category, we have seen that presheaves provide a natural semantics for
quantum truth.  Context-dependent value assignments form presheaves on
the category of contexts; the failure of a global section becomes a
precise expression of the breakdown of classical, context-independent
description.  The internal logic of the resulting presheaf topos is
intuitionistic rather than Boolean, and concrete schemes such as the
Ghose--Patra seven-valued logic can be understood as finite Heyting
algebras encoding distinct patterns of truth, falsity and
indeterminacy across incompatible contexts.  Sheaves then mark the
regime in which local data can be glued into global descriptions:
classical physics appears as the special case in which presheaf data
satisfy a sheaf condition and Boolean logic is effectively restored.

On this background, the traditional postulate of collapse can be
reinterpreted as a logical projection rather than a mysterious
physical process.  Bohr's insistence on classical language is seen as
a demand for sheaf-like, globally communicable descriptions, while the
EPR argument and Bell's theorem are recognised as attempts to impose a
global truth assignment where only contextual presheaf data are
available.  Schr\"odinger's unease about the coexistence of linear
evolution and its suspension during measurement is thus read not as
evidence for a physical discontinuity, but as an early recognition of
a mismatch between the formalism and the classical logical frame into
which it was being forced.

The $\sigma$--$\lambda$ dynamics provides a physical realisation of
the continuous passage between these logical regimes.  By interpreting
the quantum potential as a $\lambda$-scaled modification of the
Hamilton--Jacobi dynamics, linked to an underlying stochastic parameter
$\sigma$, one obtains a smooth interpolation from strongly contextual
quantum behaviour to approximately classical behaviour in which
cohomological obstructions vanish and global sections emerge.  The
dissolution of the measurement problem and the disappearance of
``nonlocality'' therefore coincide: both arise from abandoning the
requirement of a single global Boolean description in favour of a
contextual, presheaf-based semantics that is sheafified only where the
physical organisation of contexts permits it.

\paragraph{Quantum gravity and the sheafification of spacetime.}
A natural question is how the present perspective bears on quantum gravity.
In the canonical approach, the central condition is the Wheeler--DeWitt
constraint,
\(\widehat{\mathcal H}\Psi = 0\), understood as the quantization of the
Hamiltonian constraint of general relativity \cite{DeWitt1967}.
Although this is a manifestly quantum equation, its semiclassical
(Born--Oppenheimer/WKB) expansion provides a standard bridge to classical
spacetime: for a wavefunctional of the form
\(\Psi[h]\sim A[h]\exp(iS[h]/\hbar)\),
the leading order yields the Einstein--Hamilton--Jacobi equation for
\(S[h]\), reproducing classical Einstein dynamics for a chosen geometric
branch, with higher orders describing quantum fields on that background
\cite{Kiefer2012,KieferWichmann2018}.
From the present viewpoint, the recovery of classical general relativity may
therefore be read as a two-step emergence: first, restriction to an
appropriate quasi-classical branch within the constraint-satisfying sector;
second, restoration of \emph{descent} so that locally defined geometric data
(chart, frame or connection descriptions on overlapping regions) glue into a
globally communicable spacetime geometry.
This use of gluing as a diagnostic of ``classicality'' resonates with earlier
proposals to bring topos and sheaf ideas to quantum theory and quantum gravity
\cite{IshamButterfield2000} and with finitary sheaf approaches aimed at
approximating continuum spacetime structure \cite{Raptis2000}.
In our setting, the distinctive claim is that such restoration of global
gluing can be tied to a continuous classicalization controlled by the
\(\sigma\)--\(\lambda\) dynamics.

Several directions for further work suggest themselves.  On the
conceptual side, one may seek a more systematic classification of
contextual logics within presheaf topoi and a clearer comparison with
other approaches to quantum logic, including the Birkhoff--von Neumann
programme and various topos-theoretic reconstructions.  On the
mathematical side, it would be natural to refine the cohomological
analysis of contextuality and to explore more fully how the
$\sigma$--$\lambda$ dynamics controls the transition between
nontrivial and trivial cohomology.  On the physical side, one may
study concrete models---for instance, simple interferometric or
spin systems---in which the deformation from presheaf-like to
sheaf-like behaviour can be analysed quantitatively, and investigate
whether intermediate regimes admit experimental signatures.
Whatever the outcome of such developments, the central lesson remains:
logic is not a neutral backdrop for physics, but an emergent structure
reflecting the way our physical world organises its contexts.

\section{Appendix A: Introduction to Category Theory}
Having given an intuitive account of the main thesis being presented here, a slightly more technical 
account follows for the interested reader. Category theory is often described as the \emph{mathematics of structure and
relationships}. Instead of focusing on elements inside sets or algebraic
structures, it studies \emph{objects} and \emph{arrows (morphisms)} between
them.

A category $\mathcal{C}$ consists of:
\begin{itemize}
    \item objects $A,B,C,\dots$
    \item morphisms (arrows) $f:A\to B$
    \item composition: if $f:A\to B$ and $g:B\to C$, then 
    $g\circ f: A\to C$
    \item identity arrows: for each object $A$, an arrow 
    $\mathrm{id}_A:A\to A$
\end{itemize}
satisfying:
\begin{itemize}
    \item associativity: $(h\circ g)\circ f = h\circ (g\circ f)$
    \item unit laws: $\mathrm{id}_B\circ f = f = f\circ \mathrm{id}_A$
\end{itemize}

Despite its simplicity, this structure is powerful enough to unify algebra,
topology, logic, computation, and quantum theory.

Some basic examples are sets (objects are sets and morphisms are functions).
groups (objects are groups and morphisms are group homomorphisms),
topological spaces (objects are spaces and morphisms are continuous maps),
Hilbert spaces (objects are Hilbert spaces and morphisms are bounded linear maps) and
posets.

Category theory tells us:
\begin{itemize}
    \item what structures preserve what structures (via morphisms);
    \item how structures compose;
    \item when two structures should be considered the same (up to categorical equivalence rather than elementwise equality).
\end{itemize}

A functor $F:\mathcal{C}\to\mathcal{D}$ assigns:
\begin{itemize}
    \item to each object $A$ an object $F(A)$,
    \item to each arrow $f:A\to B$ an arrow $F(f):F(A)\to F(B)$,
\end{itemize}
preserving identities and composition.

Examples include:
\begin{itemize}
    \item the forgetful functor $\textbf{Grp}\to\textbf{Set}$,
    \item the free functor $\textbf{Set}\to\textbf{Grp}$,
    \item homology functors in algebraic topology.
\end{itemize}

\subsubsection*{Natural Transformations}

Given functors $F,G:\mathcal{C}\to\mathcal{D}$, a natural transformation
$\eta:F\Rightarrow G$ assigns to each object $A$ a morphism
$\eta_A:F(A)\to G(A)$ such that for every arrow $f:A\to B$,
\[
G(f)\circ \eta_A = \eta_B \circ F(f).
\]

Natural transformations express canonical, structure-preserving comparisons
between functors.

A construction such as a product, coproduct, limit, colimit, kernel, etc.\ is
defined not by its internal elements but by how it relates to other objects.

Universal properties ensure uniqueness up to unique isomorphism.

\subsubsection*{Adjunctions}

Functors $F:\mathcal{C}\to\mathcal{D}$ and $G:\mathcal{D}\to\mathcal{C}$ form
an adjoint pair, written $F\dashv G$, if there is a natural isomorphism:
\[
\mathrm{Hom}_{\mathcal{D}}(F(X),Y) \cong 
\mathrm{Hom}_{\mathcal{C}}(X,G(Y)).
\]

Adjunctions explain:
\begin{itemize}
    \item free/forgetful constructions,
    \item Galois connections,
    \item the categorical meaning of quantifiers,
    \item dualities and fundamental constructions in algebra and topology.
\end{itemize}

Let $\mathcal{C}$ be a category of \emph{contexts}.  
A \emph{presheaf} on $\mathcal{C}$ is a functor
\[
F:\mathcal{C}^{op} \to \mathbf{Set}.
\]

It assigns:
\begin{itemize}
    \item a set $F(C)$ of data available in context $C$;
    \item for every refinement $f:C\to D$, a restriction map
    \[
    F(f):F(D)\to F(C).
    \]
\end{itemize}

\noindent\emph{Physically:}
\begin{itemize}
    \item Context $C$ may be a measurement setting, coarse-graining scale, 
          reference frame, or $\sigma$--$\lambda$ diffusion regime.
    \item $F(C)$ is the set of observables, trajectories, or field-values 
          accessible in context $C$.
    \item $F(f)$ expresses how data in a finer context restricts to a coarser one.
\end{itemize}

Presheaves impose \emph{no global consistency}.  
They naturally model \emph{contextuality} and \emph{quantum-like behaviour}.

\subsubsection*{Sheaves: Gluing of Local Data}

A presheaf $F$ is a \emph{sheaf} if compatible data on overlapping contexts 
can be uniquely glued into a global piece of data.

Given a cover $\{U_i\}$ of $U$:
\[
s_i \in F(U_i),\quad 
s_i|_{U_i\cap U_j}=s_j|_{U_i\cap U_j}
\]
implies
\begin{itemize}
    \item existence: a global $s\in F(U)$ with $s|_{U_i}=s_i$, 
    \item uniqueness: $s$ is unique.
\end{itemize}

\noindent\emph{Physically:}
\begin{itemize}
    \item Classical fields behave as sheaves.
    \item Classical probability distributions behave as sheaves.
    \item Classical limits of $\sigma$--$\lambda$ dynamics (as $\lambda\to 0$) 
          approach sheaf-like behaviour.
\end{itemize}

\subsubsection*{Why Quantum Physics is Presheaf-Based}

Quantum systems violate the gluing condition.  
Local sections exist, but global sections do not:
\[
\Gamma(F)=\varnothing.
\]

This is the mathematical signature of \textbf{contextuality}. 
 
\subsection*{Sheafification = Classicalization}

For any presheaf $F$, there is a canonical \emph{sheafification} $F^{\sharp}$, 
which forces gluing by construction. 
\noindent\emph{Physically:}
\[
\text{presheaf} \quad\text{(quantum-like)} 
\quad\longrightarrow\quad 
\text{sheaf} \quad\text{(classical)}.
\]

\subsection*{Cohomology: What It Measures}

Given a presheaf $F$, the first \v{C}ech cohomology group
\[
\check{H}^1(\mathcal{C}, F)
\]
measures the \emph{obstruction to gluing} local sections into a global one.

\subsubsection*{A \v{C}ech-cohomological Obstruction to Contextuality}

Let $\{C_i \to C\}_{i\in I}$ be a family of (overlapping) contexts covering a given context $C$
(e.g.\ the maximal compatible measurement contexts of an experimental scenario). 
Write $C_{ij}:=C_i\cap C_j$ and $C_{ijk}:=C_i\cap C_j\cap C_k$ for pairwise and triple overlaps.

To speak of \v{C}ech cohomology one works with an \emph{abelian} coefficient presheaf.  
A convenient choice---used explicitly in cohomological contextuality---is to pass from a
Set-valued presheaf $F$ to the presheaf of free abelian groups $\mathbb{Z}[F]$ generated by its sections,
so that formal differences of restricted sections are meaningful \cite{AbramskyMansfieldBarbosa2012}.

A choice of local data is a \v{C}ech \emph{$0$-cochain}
\[
s=(s_i)_{i\in I},\qquad s_i\in F(C_i).
\]
Its coboundary is the $1$-cochain $(\delta s)=(\delta s)_{ij}$ given on overlaps by
\[
(\delta s)_{ij}\;:=\; s_j|_{C_{ij}} \; - \; s_i|_{C_{ij}}
\;\in\; \mathbb{Z}[F](C_{ij}).
\]
The sheaf \emph{gluing} condition is precisely $\delta s=0$ (pairwise agreement on overlaps),
in which case the family $\{s_i\}$ pastes to a global section $s\in F(C)$.

When $\delta s\neq 0$, the local sections fail to glue.  Since $\delta^{1}\!\circ\delta^{0}=0$,
the cochain $\delta s$ is automatically a $1$-cocycle, and its cohomology class
\[
[\delta s]\;\in\;\check{H}^{1}(\{C_i\},\mathbb{Z}[F])
\]
is the \emph{obstruction class}: if $[\delta s]\neq 0$ then there is no global section compatible with
the given local data.

In the Abramsky--Brandenburger sheaf-theoretic framework, an \emph{empirical model} is a compatible family
of probability distributions on maximal contexts, and contextuality is the non-existence of a global section
explaining these marginals \cite{AbramskyBrandenburger2011}.  Abramsky, Mansfield and Barbosa define a
\v{C}ech-cohomological obstruction class (built from an abelian presheaf derived from the support of the model)
which \emph{vanishes whenever a global section exists}; hence non-vanishing provides a robust \emph{sufficient}
witness of contextuality \cite{AbramskyMansfieldBarbosa2012,AbramskyBarbosaKishidaLalMansfield2015}.

\subsection*{Interpretation in Physics}

\subsubsection*{(a) Classical Physics}

If all local data glue globally:
\[
\check{H}^1(\mathcal{C},F)=0.
\]
Examples:
\begin{itemize}
    \item classical EM fields ($\mathbf{E}, \mathbf{B}$),
    \item classical trajectories,
    \item classical probability distributions.
\end{itemize}

\subsubsection*{(b) Quantum or Contextual Physics}

If gluing fails:
\[
\check{H}^1(\mathcal{C},F)\neq 0.
\]

This signals:
\begin{itemize}
    \item contextuality,
    \item nonclassical phase structure,
    \item interference phenomena,
    \item nonexistence of global truth-values.
\end{itemize}

An example is the Kochen-Specker obstruction \cite{IshamButterfield1998,Johnstone2002}.

\section{Appendix B: A Seven-valued Contextual Logic}

Ghose and Patra \cite{GhosePatra2024} have proposed a many-valued and contextual logic.  In the GP scheme, the seven 
values are understood as describing how
the truth of a proposition $P$ varies across \emph{incompatible
measurement contexts}.  For example, the value corresponding to
``true-and-false'' does not assert that $P$
is both true and false in the same context; rather, it asserts that $P$
is true in some context $C_1$ and false in another, incompatible context
$C_2$.  The combined values record patterns of variation across
contexts. 

Using the quantifier $\forall$, GP have shown that the basic modes can be formally written as
(i) $\forall x\, [\phi(x) \rightarrow p(x)]$; (ii) $\forall x\, [\phi(x) \rightarrow \neg p(x)]$; (iii) $\forall x\, [\phi(x) \rightarrow q(x)]$, $x$ standing for a variable (a placeholder) which ranges over the domain of a system (like pots), $\phi$ for a well formed formula that specifies some condition (like for example `baked'), $p$ for some predicate (like say `red') and $q$ for the predicate {\em avaktavyam}. An `example' of the first of these three in plain English would be: for all $x$ (say  clay pots) the condition $\phi(x)$ (say `baked') implies that the pot is red. 

The other four compounds may be written as 

(iv) $\forall x\,[\phi(x) \rightarrow p(x) \land \phi^\prime(x)\rightarrow \neg p(x)] \land \neg [ \phi(x) \leftrightarrow \phi^\prime (x)]$,  

(v) $\forall x\,[\phi(x) \rightarrow p(x) \land \phi^\prime(x)\rightarrow q(x)] \land \neg [ \phi(x) \leftrightarrow \phi^\prime (x)]$,  

(vi) $\forall x\,[\phi(x) \rightarrow \neg p(x) \land \phi^\prime(x)\rightarrow q(x)] \land \neg [ \phi(x) \leftrightarrow \phi^\prime (x)]$,  
 
(vii) $\forall x\,[\phi(x) \rightarrow p(x) \land \phi^\prime(x)\rightarrow \neg p(x) \land \phi^{\prime\prime}(x) \rightarrow q(x)] \land \neg [\phi(x) \leftrightarrow \phi^\prime (x)] \land\neg[ \phi^\prime (x) \leftrightarrow \phi^{\prime\prime}(x)] \land \neg[\phi(x)\leftrightarrow \phi^{\prime\prime}(x)]$.
  
Written in this formal way, the seven predications are self-consistent as they hold under {\em mutually exclusive} conditions.

\section*{Acknowledgements}

The author acknowledges the use of ChatGPT (OpenAI) for language polishing and assistance with \LaTeX{} formatting. The author takes full responsibility for the scientific content, interpretations, and any remaining errors.

\end{document}